# Edge shape sensation presented in a noncontact manner using airborne ultrasound

Koichi Kato[1], Tao Morisaki[2], Shun Suzuki[1], Yasutoshi Makino[1] and Hiroyuki Shinoda[1]

[1] *Graduate School of Frontier Sciences, The University of Tokyo, Japan*

[2] *NTT Communication Science Laboratories, Nippon Telegraph and Telephone Corporation, Atsugi, Japan*

(Email: kato@hapis.k.u-tokyo.ac.jp)

**Abstract ---** To perceive 3D shapes such as pyramids, the perception of planes and edges as tactile sensations is an essential component. This is difficult to perceive with the conventional vibrotactile sensation used in ultrasound haptics because of its low spatial resolution. Recently, it has become possible to produce a high-resolution pressure sensation using airborne ultrasound. By using this pressure sensation, it is now possible to reproduce a linear, sharp-edged sensation in the area of a fingerpad. In this study, it is demonstrated that this pressure sensation can be used to reproduce the feeling of fine, sharp edges, and its effectiveness is confirmed by comparing it with conventional vibrotactile sensation. In the demonstration, participants can experience the contact sensation of several types of edges with different curvatures.

**Keywords:** ultrasound haptics, virtual reality, edge sensation

## 1 Introduction

Tactile presentation technology is used in entertainment applications such as VR, games, and movies to enhance immersion in visual content, etc. Midair haptics technology using Airborne Ultrasound Tactile Display (AUTD) can present tactile stimulus without the need to wear a device [1, 2, 3]. AUTDs are devices with an array of ultrasound transducers. By controlling the phase of each transducer and focusing the ultrasound waves, AUTD can present a non-contact force, called acoustic radiation pressure, at an arbitrary point in the air. AUTD has been applied to various research projects, including midair touch panels [4, 5] and hand guidance [6, 7].

Since the radiation force is too weak (several tens mN) for humans to perceive, modulation of the presented radiation force distribution is needed to evoke tactile stimuli. As a modulation method, amplitude modulation (AM) and lateral modulation (LM) have been proposed. In the case of AM, the radiation force at the presented ultrasound focus is periodically changed. In the case of LM, a presented position of the focus is perceptively moved along with the skin surface. With these modulation techniques, the radiation force distribution on human skin is always dynamically varied. Since this force variation, the evoked tactile sensation by focused ultrasound has been limited to vibration or movement sensation.

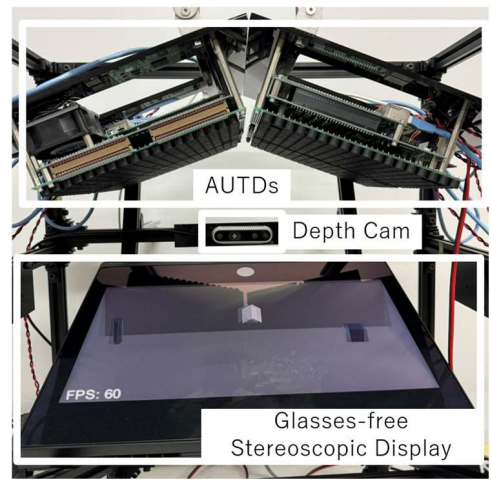

Fig.1 Ultrasonic haptics setup used in this demonstration

Recently, Morisaki et al. found that low-frequency (several Hz) LM stimuli can evoke static pressure sensation rather than vibration or movement sensation. Morisaki et al. presented a circular focus movement at 5 Hz and found that it can evoke a pressure sensation of 0.24 N on a finger pad. A pressure sensation is essential for the tactile reproduction of real objects because this sensation is always perceived when touching real objects. Moreover, since pressure sensation is perceived with higher spatial resolution than vibration, the pressure sensation by ultrasound can be used to reproduce tactile surface texture in midair [8].

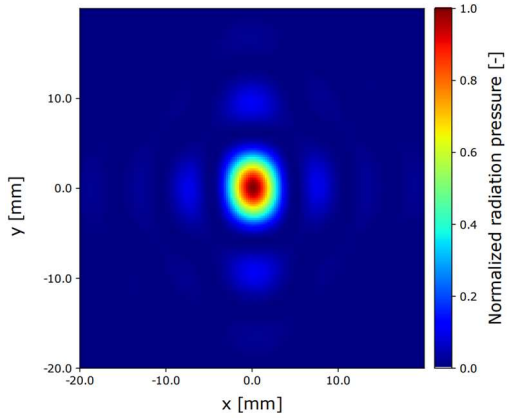

Fig.2 Sound pressure simulation of the ultrasound focus in the setup used in this demonstration

In this study, using the pressure sensation perceived with high spatial resolution, we demonstrate a sharp static edge sensation can be presented in midair. Visitors can see a midair 3D pyramid image and freely touch its top edge. When touching the edge of the image, a linear focal trajectory at 5 Hz is presented on the finger pad, evoking a sharp edge tactile sensation. We also presented a rounded edge sensation by enlarging the focal trajectory (i.e., modifying the trajectory from line to ellipse).

## 2 MID-AIR HAPTICS SYSTEM COMBINED WITH GLASSES-FREE STEREOSCOPIC IMAGING

As shown in Fig. 1, the system consists of four AUTD devices [9], a glasses-free stereoscopic display (ELF-SR1 Special Reality Display; Sony), and a depth camera (RealSense D435; Intel) for sensing 3D information on the fingerpad, The AUTD is tilted at a 20-degree angle to efficiently focus acoustic energy on the fingerpad. The emitted ultrasound waves are reflected by the ELF-SR1 to combine the 3D image with tactile stimuli.

When a participant touches a virtual object floating in front of the display, an ultrasound tactile stimulus is presented to the touch position. The touch position is detected by the depth camera.

Fig. 2 shows the simulated radiation pressure distribution of an ultrasound focus presented to the participant's fingerpad. The diameter of the ultrasound focus was about 4 mm (FWHM) in the x-direction.

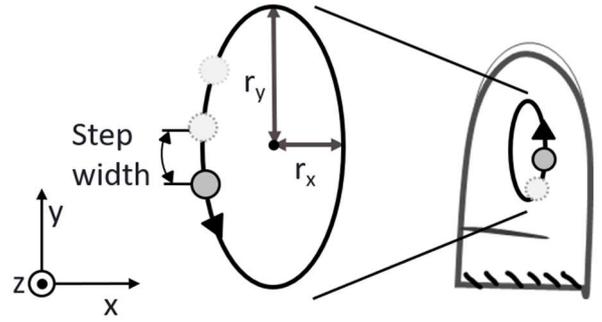

Fig.3 the schematic illustration of the stimulus pattern

For this simulation, the transducers were assumed to be point sources, and the directivity of the sound was based on the datasheet. Additionally, the radiation pressure distribution was assumed to be proportional to the square of the sound pressure.

## 3 STIMULUS PATTERN PRESENTED IN THE DEMONSTRATION

Fig. 3 shows the schematic illustration of the stimulus pattern used in the demonstration and its coordinate system. A finger is along the y-axis in this figure.

In this demonstration, we use an elliptical focal trajectory to present a sharp or rounded edge sensation. In the stimulus, the focal point passes through a point on the trajectory at the LM frequency of 5 Hz. The x- and y-axis radii of the ellipse are defined as $r_x$ and $r_y$, respectively. The stimulus presented as the sharpest edge in the demonstration has $r_x = 0$ mm, i.e., the trajectory is completely linear. The spatial step width of the focal shift is as fine as 0.2 mm and was used to produce a pressure sensation [10]. An elliptical focal trajectory with $r_x = 1$ mm was also prepared to experience touching a rounded edge.

## 4 CONCLUSION

In this study, we proposed the reproduction of sharp-edge tactile sensation in midair using pressure sensation presentation by linearly moving ultrasound focus. In the future, we will quantitatively evaluate the reality of the edge sensation.